\documentclass[]{spie}  

\usepackage{amsmath,amsfonts,amssymb}
\usepackage{graphicx}
\usepackage[colorlinks=true, allcolors=blue]{hyperref}
\usepackage{tikz}     
\usepackage{caption}
\usepackage{subcaption}
\usepackage{lipsum} 
\usepackage{xcolor}
\usepackage{array}

\title{AI-assisted prostate cancer detection and localisation on biparametric MR by classifying radiologist-positives}

\author[a]{Xiangcen Wu}
\author[a]{Yipei Wang}
\author[a]{Qianye Yang}
\author[b]{Natasha Thorley}
\author[b]{Shonit Punwani}
\author[b]{Veeru Kasivisvanathan}
\author[a,c]{Ester Bonmati}
\author[a]{Yipeng Hu}
\affil[a]{Centre for Medical Image Computing, Department of Medical Physics and Biomedical Engineering, University College London, London, UK}
\affil[b]{Div of Surgery \& Interventional Sci, University College London, London, UK}
\affil[c]{School of Computer Science and Engineering, University of Westminster, London, UK}

\begin{document} 
\maketitle

\begin{abstract}
Prostate cancer diagnosis through MR imaging have currently relied on radiologists' interpretation, whilst modern AI-based methods have been developed to detect clinically significant cancers independent of radiologists. In this study, we propose to develop deep learning models that improve the overall cancer diagnostic accuracy, by classifying radiologist-identified patients or lesions (i.e. radiologist-positives), as opposed to the existing models that are trained to discriminate over all patients. We develop a single voxel-level classification model, with a simple percentage threshold to determine positive cases, at levels of lesions, Barzell-zones and patients. Based on the presented experiments from two clinical data sets, consisting of histopathology-labelled MR images from more than 800 and 500 patients in the respective UCLA and UCL PROMIS studies, we show that the proposed strategy can improve the diagnostic accuracy, by augmenting the radiologist reading of the MR imaging. Among varying definition of clinical significance, the proposed strategy, for example, achieved a specificity of 44.1\% (with AI assistance) from 36.3\% (by radiologists alone), at a controlled sensitivity of 80.0\% on the publicly available UCLA data set. This provides measurable clinical values in a range of applications such as reducing unnecessary biopsies, lowering cost in cancer screening and quantifying risk in therapies.
\end{abstract}


\section{INTRODUCTION}
\label{sec:intro} 
Multi-parametric MR (mpMR) imaging has been developed for a non-invasive alternative to previous blind biopsies, for detecting and localise clinically significant prostate cancer (csPC). This has motivated many machine learning (ML) methods developed for this MR-based radiological task. For classifying patients with csPC from those who may safely avoid unnecessary biopsy confirmation, studies showed that ML models achieved as accurately as radiologists \cite{schelb2019classification}, or even better \cite{liu2017prostate}. However, both radiologists and ML classifiers are still subject to limited performance. For example, the PROMIS study showed a specificity of 38.9\% at a controlled sensitivity of \~90\%.
Few studies evaluated the diagnostic accuracy at local level, such as localising individual lesions or Barzell zones \cite{Barzell, Identifying} - by radiologists or ML models despite its clinical utilities in targeted biopsy, therapy planning and evaluation. This is in part due to a lack of data with comprehensive histopathology labels, for robust validation. Efforts have been made, however, through the use of template-based saturation biopsy, e.g. in the PROMIS study \cite{AHMED2017815}, and a mixed targeted and systematic biopsy \cite{Natarajan2020}, to provide better sampling of the disease.


The existing ML models, which have access to both radiologist and histopathology labels during training, has the potential to improve the radiologist-only performance \cite{zeevi2023reliable}. Histopathology labels obtained from a mixed targeted and systematic biopsy (e.g. the UCLA data set \cite{Natarajan2020} used in this study) are more feasible, therefore more available, compared with saturation biopsy. However, radiologist-negative regions are sampled randomly and highly sparsely (e.g. 3 - 30 needle biopsied specimens) in the former. This directly leads to a highly variable and, potentially, unknown biased histopathology labels (biased due to, e.g., cohort, observer, imaging protocol). Our preliminary experience suggested that training ML models learned from such labels can be challenging and inefficient. In contrast, the radiologist-positive cases are sampled by targeted biopsy which yields lower (albeit perhaps still significant) variance and bias, which motivated this work. 

In this work, we aim to develop an AI system that provides assistance to radiologist MR reading, rather than performing independently of the radiologists. We propose to train the AI models solely on classifying radiologist-positive cases (at lesion, Barzell zone and patient levels), to avoid using the ``radiologist-negative cases'' with the above-discussed label issues. We argue that those radiologist-positive cases have 1) a better-defined population distribution (based on continuously improving radiology guidelines) and 2) better sampled histopathology labels (e.g. through targeted biopsy), compared to the ``entire'' population (i.e. with radiologist-negative cases) often with varying entry criteria and lacking consistent pathology labels from often under-sampled patients or regions. The developed classifiers then can be combined with the radiologists to provide the final and improved diagnosis.

\begin{figure}[htbp]
    \centering
    \subfloat[]{
        \includegraphics[width=0.25\textwidth]{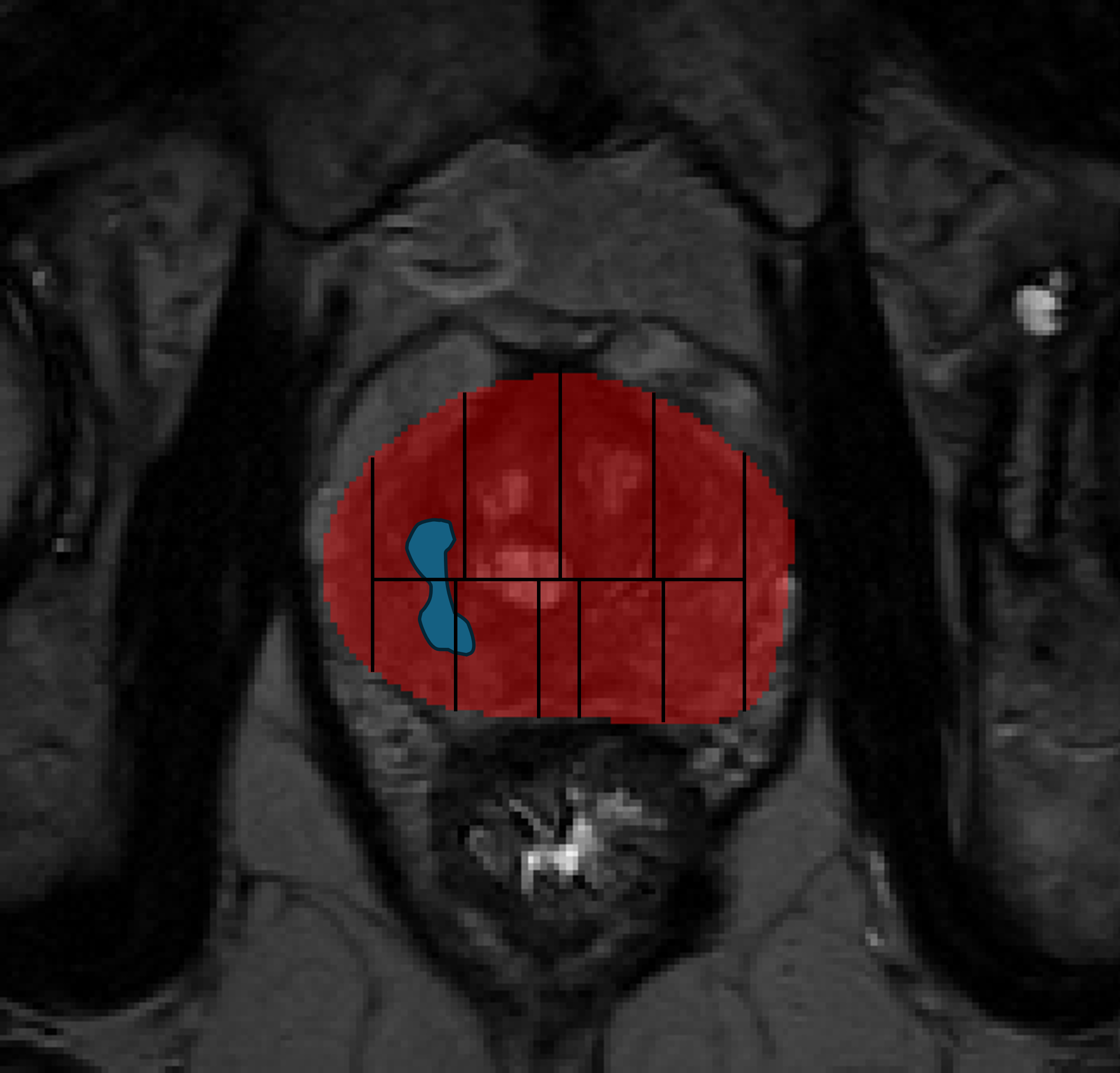}
        \label{BZoneConvert}
    }
    \hfill
    \subfloat[]{
        \includegraphics[width=0.7\textwidth]{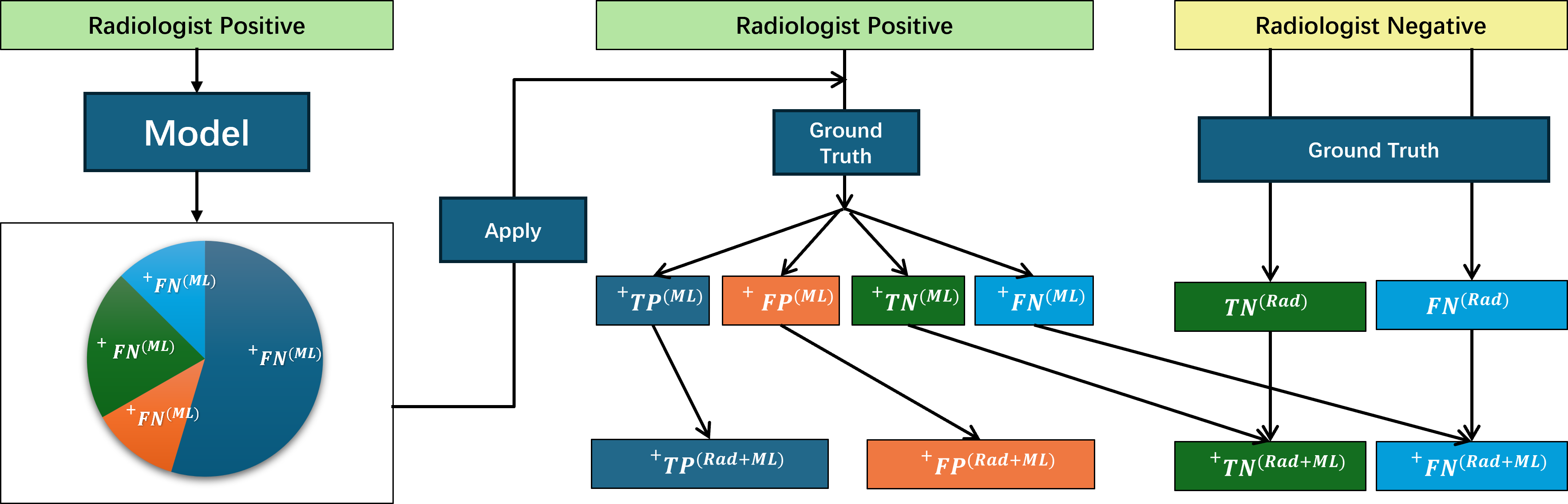}
        \label{Zone level validation workflow}
    }
    \caption{(a) A suspicious lesion covering three zones based on Barzell zone template. (b) A diagram illustrates the proposed classification of radiologist-positive cases.}
    
\end{figure}
\section{Method}

\subsection{Classification of radiologist-positive cases}
\label{sec:method-classificiation}
Training an ML model to classify MR voxels, histopathology ground-truth is required at every voxel, to compute the number of true positive $TP^{(ML)}$, false positive $FP^{(ML)}$, true negative $TN^{(ML)}$ and false negative $FN^{(ML)}$ cases. Similarly, the radiologist performance can be measured by $TP^{(Rad)}$, $FP^{(Rad)}$, $TN^{(Rad)}$ and $FN^{(Rad)}$. In targeted biopsy, annotation from radiologists indicates suspicious regions of interest (ROIs) which undergo needle-based biopsy sampling for subsequent histopathology examination. Given a definition of clinically significant cancer (e.g. Gleason group $\geq$ 3+4), our proposed ML models classify those radiologist-positive ROIs ($^{+}$ROIs) into $^{+}TP^{(ML)}$, $^{+}FP^{(ML)}$, $^{+}TN^{(ML)}$ and $^{+}FN^{(ML)}$, based on positive/negative pathology. The prefix $^{+}$ indicates the classification of radiologist-positive cases, where the ``cases'' refers to radiologist's annotation on MR imaging voxels during model training. The voxels outside of these $^{+}$ROIs are, by definition, considered as negative (or a separate background class) labels during training, therefore their histopathology labels are not required. During model evaluation, these classified voxels are subsequently grouped and analysed at different sampling levels (i.e. now ``cases'' are lesions, Barzell zones\cite{Barzell} or patients, based on available ground-truth data). The following section describes our proposed implementation, using the radiologist-positive cancer classification system, to classify all cases into $TP^{(Rad+ML)}$, $FP^{(Rad+ML)}$, $TN^{(Rad+ML)}$ and $FN^{(Rad+ML)}$, indicated in Figure~\ref{Zone level validation workflow}. It is noteworthy that $ ^{+}TP^{(ML)} = TP^{(Rad+ML)}$ and $^{+}FP^{(ML)} = FP^{(Rad+ML)}$.

\subsection{Segmentation networks}
\label{sec:method-network}
\begin{figure}[htbp]
    \centering
    \subfloat[Radiologist true (right) and false (left) positive]{
        \includegraphics[width=0.5\textwidth]{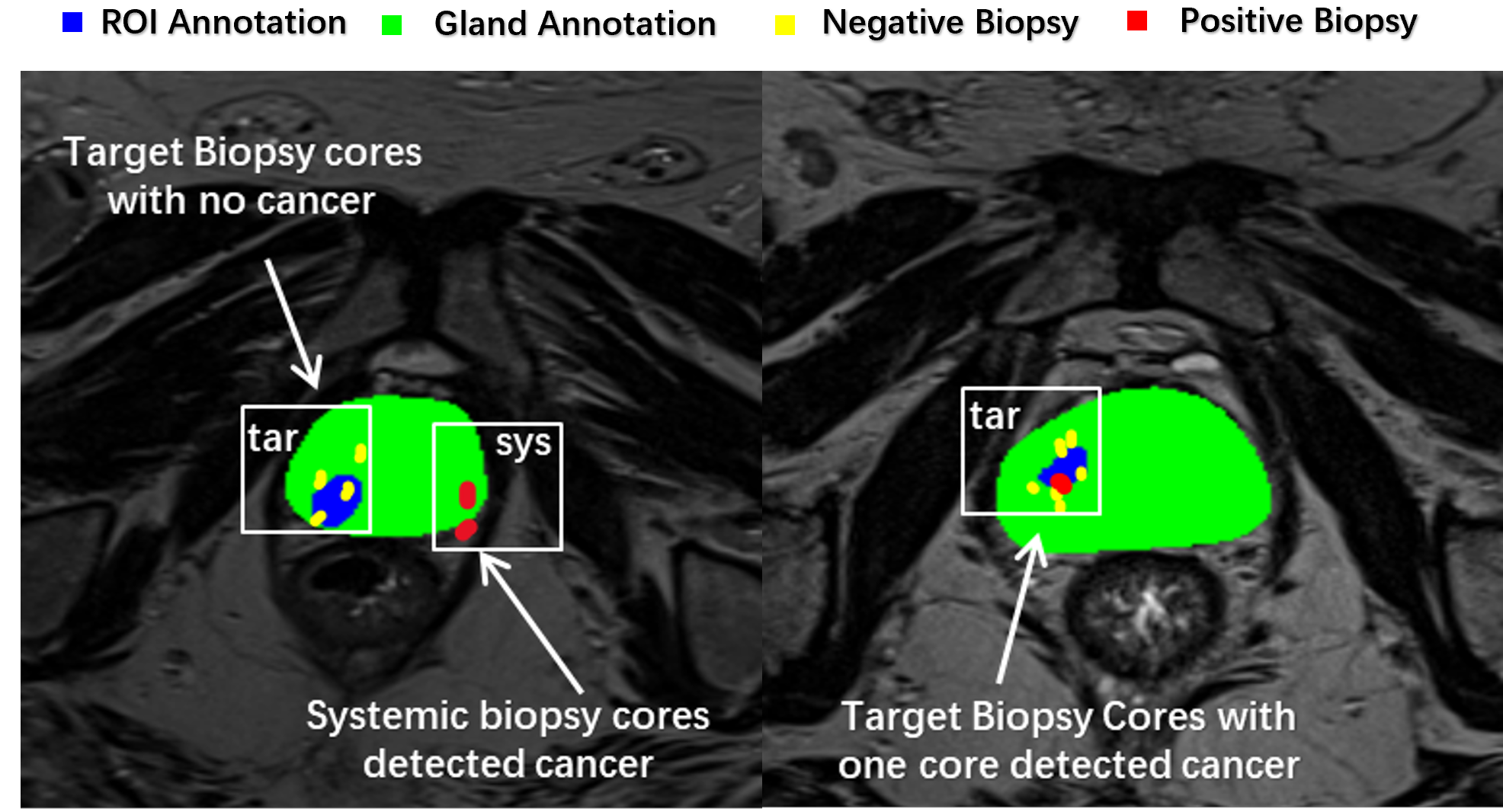}
        \label{fig:TPFP Determination}
        
    }
    \hfill
    \subfloat[Model inference of radiologist's annotations]{
        \includegraphics[width=0.45\textwidth]{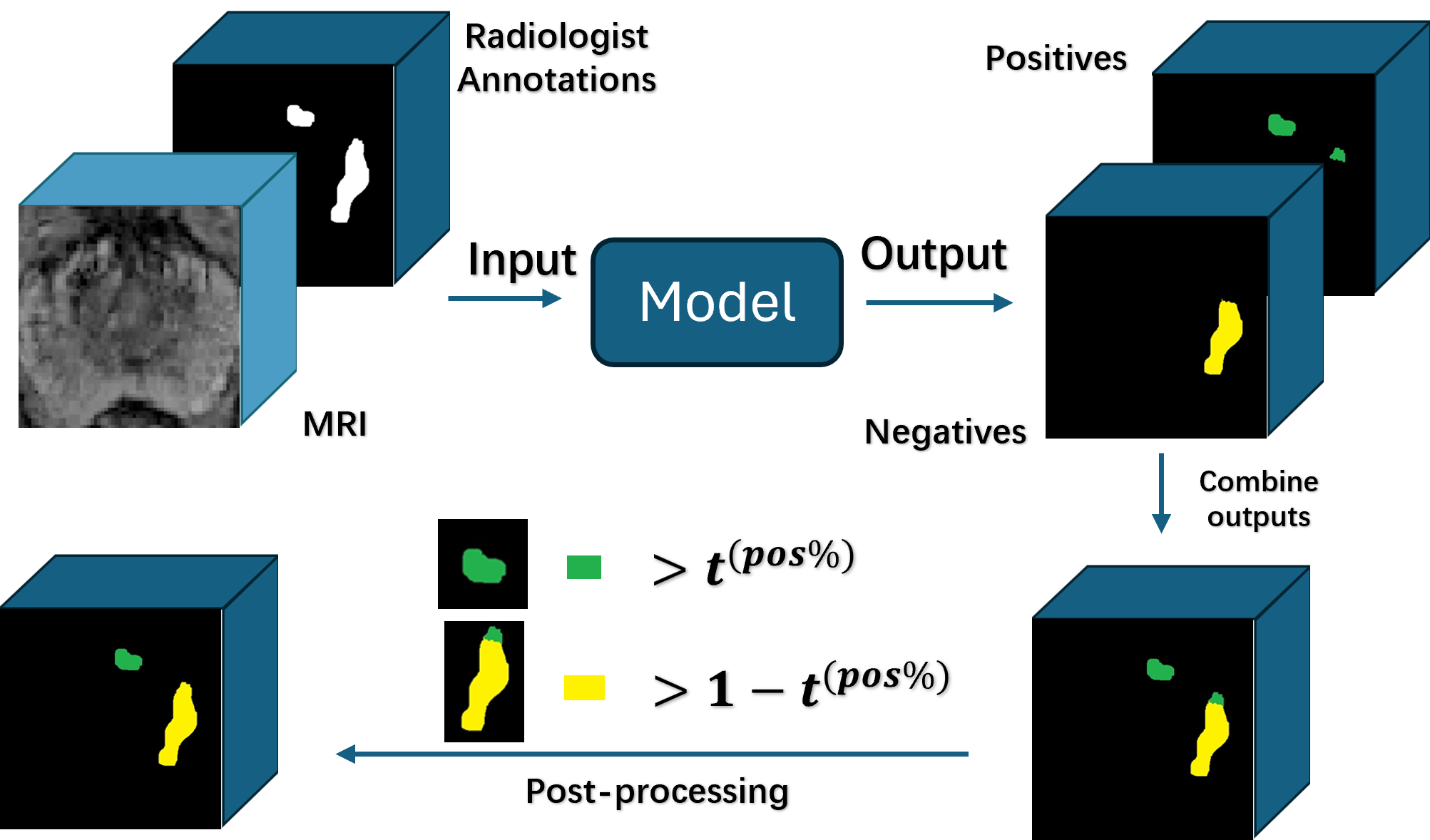}
        \label{model workflow}
    }
    \caption{Determination of suspicious ROIs on showed on (a), and (b) structural diagram of the deep learning prediction and classification of positive/negative on radiologist's annotations.}
    \label{ml-workflow}
\end{figure}

A segmentation network is adopted to take MR images and the radiologist annotation as the multi-channel network input. In this work, a two-channel input contains a T2-weighted (T2w) MR volume and corresponding radiologist-positive segmentation mask, and a four-channel bi-parametric MR (bpMR) input uses a T2w, a diffusion-weighted with high b-value (DWI$_{hb}$), an ADC map and a segmentation mask. The input may be configured to take further imaging modalities or, potentially, other clinical data if available. Although mpMR imaging is now widely accepted for accurate prostate detection, testing the case of using uni-modal T2w-only input is interesting. This is because classifying radiologist-positive cases has not been previously explored, thus the sensitivity of the learned model discriminating ability to different modalities is yet investigated.
The segmentation model is trained to predict a 3-class class probability map, i.e. with three output channels representing the positive, negative and background classes, supervised by the radiologist-positive labels, described in Secs.~\ref{sec:method-classificiation} and~\ref{sec:exp-levels}. During inference, a threshold is used in this work, for adjustment between Type I and Type II errors in subsequent classification tasks. The threshold $t^{(pos\%)}\in [0,1]$ indicates the percentage of positive voxels in a ROI, over which the radiologist-defined ROIs are classified as positive/negative ROIs, as illustrated in Fig.~\ref{model workflow}.
\section{Experiments}

\subsection{Data sets labelled by varying biopsy protocols and network training}
The open UCLA data set \cite{Natarajan2020} are used primarily for model training, with T2w, ADC, and DWI$_{hb}$ sequences. 898 and 825 imaging studies, for respective T2w-only and bpMR, are annotated with both radiologist ROIs (with 0-5 UCLA scores) and histopathology ground-truth. The ground-truth is based on a mixed targeted biopsy and systematic random biopsy outside of the targeted ROIs. All image volumes are center-cropped based on the radiologist annotation of the prostate gland, intensity-normalized, and resampled to an image size of $64\times64\times64$. The recently published PROMIS data set \cite{AHMED2017815} are used in an additional validation of the radiologist-and-ML combined performance, in part due to its more densely sampled histopathology results outside of the radiologist-positive ROIs (with 1-5 PIRADS scores), based on Barzell zones \cite{Barzell}. 
 

We utilize SwinUNETR-v2 \cite{swin-v2} as our segmentation model. All results reported in this study are based on multiple random train-test splits, repeated for five times. For each split, we train the model for a fixed 100 epochs using the AdamW optimizer with a learning rate of $5e^{-5}$. During training, we apply random affine transformations, Gaussian smoothing, Gaussian noise, and contrast adjustment, each with a probability of 0.25.

\subsection{Patient, ROI and Barzell zone classification}
\label{sec:exp-levels}
Patient-level accuracy can be computed on the UCLA data set, with positive patients being indicated by any positive biopsy, tested at different UCLA score cutoffs $\in[2,5]$. By varying the threshold $t^{(pos\%)}$, the radiologist-and-ML combined sensitivity $sen^{(Rad+ML)}$ and specificity $spc^{(Rad+ML)}$ are compared with the radiologist performance $sen^{(Rad)}$ and $spc^{(Rad)}$. 
For classifying the radiologist-positive ROIs, the ML model performance, $^{+}sen^{(ML)}$ and $^{+}spc^{(Rad)}$, is based on $^{+}TP^{(ML)}$, $^{+}FP^{(ML)}$, $^{+}TN^{(ML)}$ and $^{+}FN^{(ML)}$ (Sec.~\ref{sec:method-classificiation}). For assessing localisation performance taking into account radiologist-negative cases, we classify Barzell zones into different csPC definitions as defined in the PROMIS study, with varying PIRADS cutoffs, to compare the resulting radiologist-and-ML combined $sen^{(Rad+ML)}$ and $spc^{(Rad+ML)}$ and the radiologist performance $sen^{(Rad)}$ and $spc^{(Rad)}$. The radiologist performance can be obtained from the PROMIS data set, based on the Barzell-zone-level $TP^{(Rad)}$, $FP^{(Rad)}$, $TN^{(Rad)}$ and $FN^{(Rad)}$. The developed ML model is applied on the radiologist-positive Barzell zones, see Fig~\ref{BZoneConvert}, based on the same approach described in Sec.~\ref{sec:method-network}, on the UCLA test data. This obtained radiologist-positive zone classification is then assumed on the PROMIS data set to modify the radiologist classification, to obtain a new set of $TP^{(Rad+ML)}$, $FP^{(Rad+ML)}$, $TN^{(Rad+ML)}$ and $FN^{(Rad+ML)}$, as described in Sec.~\ref{sec:method-classificiation}.

\section{Results and discussion}
The ROI-level, zone-level and patient-level results are summarised in Table.~\ref{table:combinedtables}. The results are reported as example sensitivity and specificity, at varying diagnostic cutoffs and csPC definitions, for assessing their respective clinical values and a direct comparison between methods, as detailed in Sec.~\ref{sec:exp-levels}. The overall accuracy metrics, such as average precision and area under ROC curves, are omitted due to their unclear clinical relevance. As seen from Table.~\ref{table:combinedtables}a, the high $^{+}sen$ and $^{+}spc$ in classifying $^{+}$ROIs are obtained at varying radiologist score cutoffs, for both T2w-only and bpMR cases, with visual examples provided in Fig.~\ref{fig:model-prediction}. For classifying Barzell zones, a $spc^{(Rad+ML)}$ of $\>$90\% was obtained by adding the proposed ML models, improved from the radiologists' $spc^{(Rad)}$ of 72.5\%, at the same $sen^{(Rad+ML)} = sen^{(Rad)}$, as in Table.~\ref{table:combinedtables}b. The improvement was also readily observed at the patient classification, as shown in Table.~\ref{table:combinedtables}c. 

Our work demonstrate a wide applicability of the proposed radiologist-positive classification approaches, for both patient risk stratification and csPC localisation, as well as in different clinical scenarios, such as screening that requires high sensitivity and managing high-risk procedures that may benefit from high specificity.

\begin{figure}[h]
    \centering
    \begin{tikzpicture}
        \node[anchor=south west,inner sep=0] (image) at (0,0) {\includegraphics[width=1\textwidth]{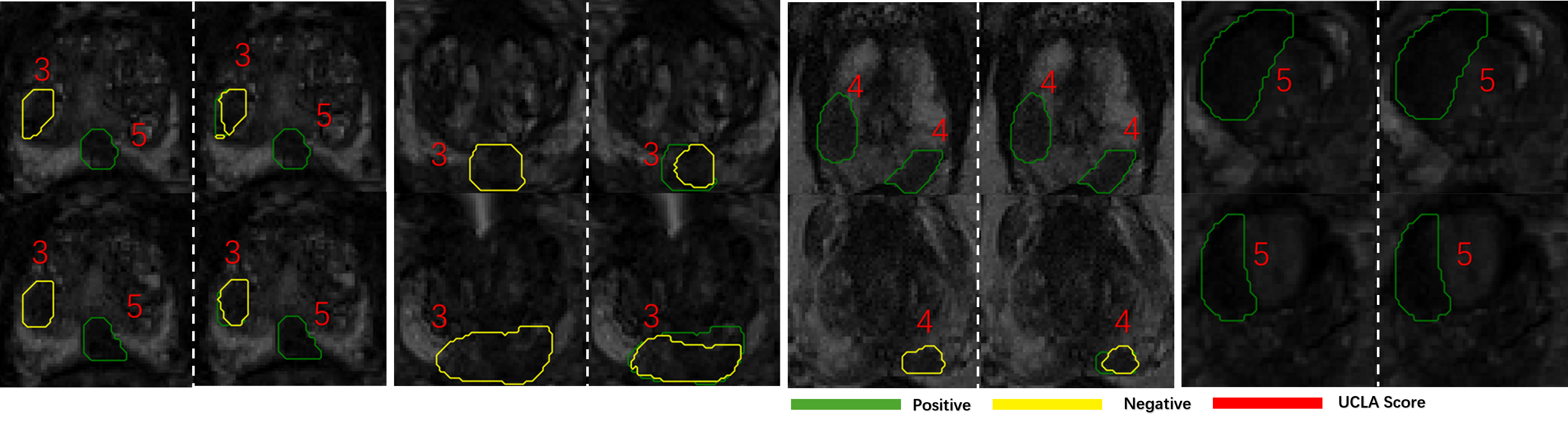}};
        
        \begin{scope}[x={(image.south east)},y={(image.north west)}]
            \node at (0.12,1.05) [text=black, font=\tiny] {Patient A};
            \node at (0.37,1.05) [text=black, font=\tiny] {Patient B};
            \node at (0.62,1.05) [text=black, font=\tiny] {Patient C};
            \node at (0.87,1.05) [text=black, font=\tiny] {Patient D};
        \end{scope}
    \end{tikzpicture}
    \caption{Four example patients were utilized here on the figure. Left side of the dashed line represents the ground truth data, while the right side displays the model's predictions.}
    \label{fig:model-prediction}
\end{figure}
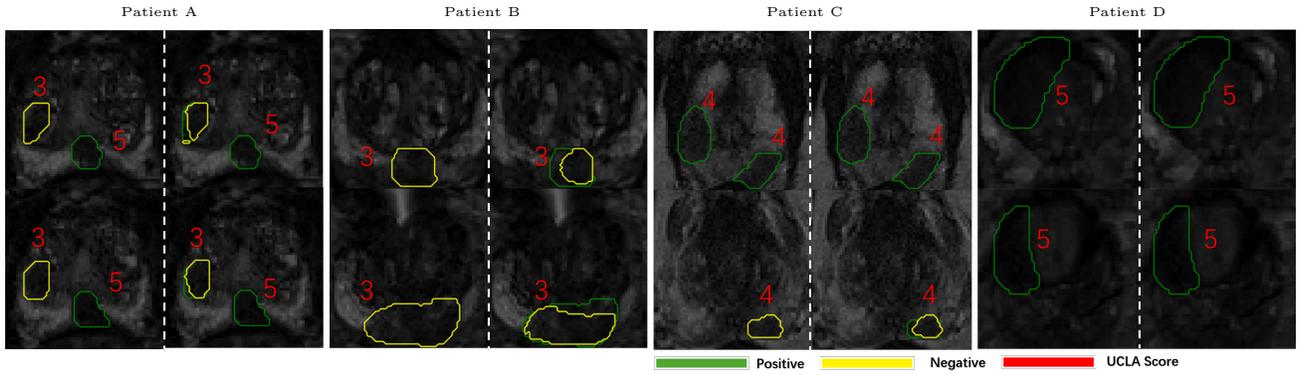

\begin{figure}[htbp]
    \centering
    \begin{minipage}[t]{0.45\textwidth}
    \vspace{-5.35cm}
    \hspace{0.8cm}
        \subfloat[ROI level results on $^{+}$ROIs]{%
            \begin{tabular}{|l|c|c|}
                \hline
                
                \textbf{}                        & \textbf{Sen} & \textbf{Spe} \\
                \hline
                \textbf{UCLA $\geq$3 (T2)}    &     94.20\%  &    97.14\%   \\ 
                \hline
                \textbf{UCLA $\geq$5 (T2)}    &      92.59\% &     76.42\%  \\ 
                \hline
                \textbf{UCLA $\geq$3 (bpMR)}    & 76.92\%      & 95.41\%      \\ \hline
                \textbf{UCLA $\geq$5 (bpMR)}    & 91.30\%      & 52.27\%      \\ \hline
   
            \end{tabular}
            \label{table:tumorlevel}
        }
        
        \subfloat[Zone level results on PROMIS]{%
        \hspace{0.3cm}
            \begin{tabular}{|c|c|c|}
                \hline
                & \textbf{Sen} & \textbf{Spe} \\
                \hline
                \textbf{Promise Radiologist} &  \textcolor{red}{58.66\%} &  \textcolor{red}{72.53\%} \\
                \hline
                \textbf{\boldmath$t^{(pos\%)}$ = 0.5 (T2)} & 38.20\% & 92.18\% \\
                \hline
                \textbf{\boldmath$t^{(pos\%)}$ = 0.3 (T2)} & 39.29\% & 91.80\% \\
                \hline
                \textbf{\boldmath$t^{(pos\%)}$ = 0.01 (T2)} & 45.69\% & 91.10\% \\
                \hline
                \textbf{\boldmath$t^{(pos\%)} =$  NaN (T2)} & \textcolor{red}{58.66\%} & \textcolor{red}{\textbf{90.42\%}} \\
                \hline
                \textbf{\boldmath$t^{(pos\%)}$ = 0.5 (bpMR)} & 30.09\% & 93.10\% \\
                \hline
                \textbf{\boldmath$t^{(pos\%)}$ = 0.3 (bpMR)} & 34.11\% & 92.58\% \\
                \hline
                \textbf{\boldmath$t^{(pos\%)}$ = 0.01 (bpMR)} & 37.39\% & 90.56\% \\
                \hline
                \textbf{\boldmath$t^{(pos\%)} =$  NaN (bpMR)} & \textcolor{red}{58.66\%} & \textcolor{red}{\textbf{75.70\%}} \\
                \hline
            \end{tabular}
            \label{fig:zonelevel}
        }
    \end{minipage}
    \hfill
    \begin{minipage}[t]{0.44\textwidth}
        \subfloat[Patient-level classification was performed using various UCLA scores as thresholds, both before and after AI assistance. The low specificity observed among radiologists (notated as $^{*}$) is anticipated due to the availability of labeled UCLA data. The ground truth of the UCLA data is biased towards positive patients.]{%
            \begin{tabular}{|c|c|c|}
                \hline
                & \textbf{Sen} & \textbf{Spe} \\
                \hline
                \textbf{$^{*}$UCLA $\geq$ 2} & 92.24\% & 6.44\% \\
                \hline
                \textbf{$^{*}$UCLA $\geq$ 3} & 92.19\% & 7.84\% \\
                \hline
                \textbf{$^{*}$UCLA $\geq$ NaN} & \textcolor{red}{80.00\%} & \textcolor{red}{36.33\%} \\
                \hline
                \textbf{$^{*}$UCLA $\geq$ 4} & 65.47\% & 70.30\% \\
                \hline
                \textbf{$^{*}$UCLA $\geq$ 5} & 29.78\% & 94.06\% \\
                \hline
                \textbf{\boldmath$t^{(pos\%)}$ = 5 (T2)} & 83.85\% & 42.86\% \\
                \hline
                \textbf{\boldmath$t^{(pos\%)}$ = NaN (T2)} & \textcolor{red}{80\%} & \textbf{\textcolor{red}{41.98\%}} \\
                \hline
                \textbf{\boldmath$t^{(pos\%)}$ = 0.9 (T2)} & 78.88\% & 41.73\% \\
                \hline

                \textbf{\boldmath$t^{(pos\%)}$ = 0.5 (bpMR)} &74.58\% & \textbf{55.56\%} \\
                \hline
                \textbf{\boldmath$t^{(pos\%)}$ = NaN (bpMR)} & \textcolor{red}{80\%} & \textcolor{red}{\textbf{44.13\%}} \\
                \hline
                \textbf{\boldmath$t^{(pos\%)}$ = 0.1 (bpMR)} & 81.3\% & 44.44\% \\
                \hline

            \end{tabular}
            \label{fig:patientlevel}
        }
    \end{minipage}
    \captionsetup{type=table,name=Table}
    \caption{Different levels of classification. NaN indicates the values are computed though linear interpolation. $^{+}$ROIs indicates the radiologist-positive cases or lesions}
    \label{table:combinedtables}
\end{figure}



\bibliography{references.bib} 
\bibliographystyle{spiebib} 

\end{document}